\documentclass[12pt]{article}
\usepackage{amsfonts,amsmath,latexsym,amssymb,makeidx,txfonts} 
\usepackage[american]{babel}

\makeatletter
\renewcommand{\@evenhead}{\raisebox{0pt}[\headheight][0pt]{\vbox{\hbox
to \textwidth{\thepage\hfil\strut\textsc{\leftmark}}\hrule}}}
\renewcommand{\@oddhead}{\raisebox{0pt}[\headheight][0pt]{\vbox{\hbox
to \textwidth{\textsc{\rightmark}\hfil\strut\thepage}\hrule}}}
\makeatother

\newcommand\II{{\mathbb I}}
\newcommand\RR{{\mathbb R}}
\newcommand\CC{{\mathbb C}}

\newcommand\tr{{\rm tr\,}}
\newcommand\Tr{{\rm Tr\,}}
\newcommand\Det{{\rm Det\,}}

\newcommand\Aut{{\rm Aut}}

\newcommand\End{{\rm End}}
\newcommand\Ind{{\rm Ind}}

\newcommand\vol{{\rm vol\,}}

\newcommand\Ker{{\rm \,Ker\,}}

\newcommand\be{\begin{equation}}
\newcommand\ee{\end{equation}}
\newcommand\bea{\begin{eqnarray}}
\newcommand\eea{\end{eqnarray}}

\def\sideremark#1{\ifvmode\leavevmode\fi\vadjust{\vbox to0pt{\vss
\hbox to 0pt{\hskip\hsize\hskip1em
\vbox{\hsize2cm\tiny\raggedright\pretolerance10000
\noindent #1\hfill}\hss}\vbox to8pt{\vfil}\vss}}}

\begin{document}

\begin{titlepage}

\null
\vskip-2cm 
\hfill
\begin{minipage}{5cm}
\par\hrulefill\par\vskip-4truemm\par\hrulefill
\par\vskip2mm\par
{{\large\sc New Mexico Tech \\[12pt]
{\rm (September 2005)}}}
\par\hrulefill\par\vskip-4truemm\par\hrulefill
\end{minipage}
\bigskip 
\bigskip

\hfill
\begin{minipage}{8cm}
to be published in:\\
Proceedings of the Workshop
{\it ``Krzysztof Wojciechowski 50 years --
Analysis and Geometry of Boundary Value Problems''},
Roskilde, Denmark, 20-22 May, 2005
\end{minipage}

\par
\bigskip
\bigskip

\centerline{\LARGE\bf Non-Laplace type Operators}
\medskip
\centerline{\LARGE\bf on Manifolds with Boundary} 
\bigskip
\bigskip 
\centerline{\Large\bf Ivan G. Avramidi} 
\bigskip 
\centerline{\it Department of Mathematics} 
\centerline{\it New Mexico Institute of Mining and Technology} 
\centerline{\it Socorro, NM 87801, USA} 
\centerline{\it E-mail: iavramid@nmt.edu} 
\bigskip
\bigskip

\medskip 
\vfill 
{\narrower 
\small
Non-Laplace type operators are second-order elliptic partial
differential operators acting on sections of vector bundles over a
manifold with boundary with a non-scalar leading symbol.  Such operators
appear, in particular, in models of non-commutative gravity theories,
when instead of a Riemannian metric there is a matrix valued
self-adjoint symmetric two-tensor that plays the role of a
``non-commutative'' metric.  We construct the corresponding heat kernel
and compute its first two spectral invariants. }

\end{titlepage}


\section{Introduction}

Elliptic differential operators on manifolds play a very important role
in mathematical physics, geometric analysis, differential geometry and
quantum theory. Of special interest are the resolvents and the spectral
functions of elliptic operators; the most important spectral functions
being the trace of the heat kernel and the zeta function, which
determine, in particular, the functional determinants of differential
operators (see \cite{gilkey95,booss93,berline92,avramidi00} and the
reviews \cite{avramidi99,avramidi02b,vassilevich03}).   

In particular, in quantum field theory and statistical physics the
resolvent determines the Green functions, the correlation functions and
the propagators of quantum fields, and the functional determinant
determines the effective action and the partition function (see, for
example \cite{avramidi00}). In spectral geometry, one is interested,
following Kac \cite{kac66}, in the question: ``Does the spectrum of the
scalar Laplacian determine the geometry of a manifold'' or, more
generally, ``{\it To what extent does the spectrum of  a differential
operator on a manifold determine the geometry of the manifold?}''  Of
course, the answer to Kac's and other questions depends on the
differential operator. Most of the studies in spectral geometry and
spectral asymptotics are restricted to so-called Laplace type operators.
These are second-order partial differential operators  acting on
sections of a vector bundle with a positive definite {\it scalar leading
symbol}. 

Since, in general, it is impossible to find the spectrum of a
differential operator exactly, one studies the asymptotic properties of
the spectrum, so-called {\it spectral asymptotics},  which are best
described by the asymptotic expansion of the trace of the heat kernel. 
If  $L: C^\infty(V)\to C^\infty(V)$ is a self-adjoint elliptic
second-order partial differential  operator with a positive definite
leading symbol acting on smooth sections of a vector bundle $V$ over a
compact  $n$-dimensional manifold $M$, then there is a small-time
asymptotic expansion as $t\to 0$ \cite{greiner71,gilkey95}
\be
\Tr_{L^2}\exp(-tL)\sim (4\pi)^{-n/2}\sum_{k=0}^\infty t^{(k-n)/2}A_k\,. 
\label{asexp}
\ee
The coefficients $A_k$ are called the global {\it heat invariants} (in
mathematical literature they are usually called the
Minakshisundaram-Pleijel coefficients; in physics literature, they are
also called HMDS (Hadamard-Minakshisundaram-De Witt-Seeley)
coefficients, or Schwinger-De Witt coefficients). 

The heat invariants  are spectral invariants of the operator $L$ that
encode the information about the asymptotic properties of the spectrum.
They are of great importance in spectral geometry and find extensive
applications in physics, where they describe renormalization and 
quantum corrections to the effective action in quantum field theory and
the thermal corrections to the high-temperature expansion in statistical
physics among many other things. They describe real physical effects 
and,  therefore, the knowledge of these coefficients in {\it explicit
closed form} is important in physics. One would like to have formulas
for some lower-order coefficients to be able to study those effects.

The proof \cite{greiner71,gilkey95} of the existence of such an 
asymptotic expansion (\ref{asexp}) has been a great achievement in
geometric analysis. Now it is a well known fact, at least in the smooth
category for compact manifolds. This is not the subject of our interest. 
The main objective in the study of spectral asymptotics (in spectral
geometry and quantum field theory) is, rather, the {\it explicit
calculation} of the heat invariants $A_k$ in {\it invariant geometric
terms}. 

The approach of Greiner and Seeley \cite{greiner71,seeley69} is  a very
powerful general analytical procedure for analyzing the structure of the
asymptotic expansion based on the theory of pseudo-differential
operators and the calculus of symbols of operators (we will call it {\it
symbolic approach} for symplicity). This approach {\it can be used} for
calculation of the heat invariants explicitly in terms of the jets of
the symbol of the operator; it provides an iterative procedure for such
a calculation. However, as far as we know,  because of the technical
complexity and, most importantly, lack of the manifest covariance, such
analytical tools {\it have never been used} for the actual calculation
of the explicit form of the heat invariants in an invariant geometric
form.  As a matter of fact, the symbolic method has only been used to
prove the existence of the asymptotic expansion and the general
structure of the heat invariants (like their dependence on the jets of
the symbol of the operator) (see the review \cite{avramidi02b} and other
articles in the same volume, and \cite{vassilevich03,kirsten01}). To the
best of our knowledge there is no exlicit formula even for the 
low-order coefficients $A_1$ and $A_2$ for a general non-Laplace type
operator.

The development of the analysis needed to discuss elliptic boundary
value problems is beyond the scope of this paper. We shall simply use
the well known results about the existence of the  heat trace
asymptotics  of elliptic boundary value problems from the classical
papers of Greiner \cite{greiner71} and Seeley \cite{seeley69} (see also
the books \cite{grubb96,booss93}). Our approach can be best described by
Greiner's own words \cite{greiner71}, pp. 165--166,: {\it ``the
asymptotic expansion can be obtained by more classical methods. Namely,
one constructs the Taylor expansion for the classical parametrix [of the
heat equation] \dots and iterates it to obtain the Green's operator.
This yields, at least  formally, the asymptotic expansion for [the trace
of the heat kernel]''}. This is the approach exploited in
\cite{mckean67} for a Laplace type operator and it is this approach that
we will use in the present paper for non-Laplace type operators. 
However, contrary to \cite{greiner71,seeley69} we do not use any
Riemannian metrics but, instead,  work directly with densities, so that
our final answers are automatically invariant. Greiner \cite{greiner71},
pp. 166, also points out that {\it ``Of course, at the moment it is not
clear which representation will yield more easily to geometric
interpretation.''}

In spectral geometry as well as in physics the motivation and the goals
of the study of spectral asymptotics are quite different from those in
analysis. The analytic works are primarily interested in the existense
and the type of the asymptotic expansion, but not necessarily in the
explicit form of the coefficients of the expansion. In spectral geometry
one is interested in the {\it explicit form} of the spectral invariants 
and their relation to geometry.  One considers various special cases
when some invariant topological and geometrical constraints are imposed,
say, on the Riemannian structure (or on the connection of a vector
bundle). Some of these conditions are: positive (negative, or zero)
scalar curvature, or positive (negative) sectional curvature, Ricci-flat
metrics, Einstein spaces, symmetric spaces, Kaehler manifolds etc. Such
conditions lead to  very specific consequences for the heat invariants
which are obvious in the geometric invariant form but which are hidden
in a non-invariant symbolic formula obtained in local coordinates. For
example, if the scalar curvature  is zero, then for the Laplacian on a
manifold without boundary  $A_2=0$. Such a conclusion cannot be reached
until one realizes that the integrand of $A_2$ is precisely the scalar
curvature. There are, of course, many more examples like this. 

Another property that does not become manifest at all in the symbolic
approach is the behavior of the heat invariants under the conformal
transformation of the Riemannian structure and the gauge
transformations. This is a very important property that is heavily used
in the functorial approach \cite{branson90,branson99}, but which is not
used at all in the symbolic approach. For conformally covariant
operators the symbolic calculus is exactly the same as for
non-conformally covariant ones with similar results because the
conformal covariance only concerns the low-order terms of the symbol but
not its leading symbol. However, the conformal invariance leads to
profound consequences for the heat invariants, zeta-function and the
functional determinant (see
\cite{branson95}).

{\it The calculation of the explicit form of the heat invariants is a
separate important and complicated problem that requires special
calculational techniques}.  The systematic explicit calculation of heat
kernel coefficients was initiated by Gilkey \cite{gilkey75} (see
\cite{gilkey95,vassilevich03,kirsten01,avramidi02b} and references
therein). A review of various algorithms for calculation of the heat kernel
coefficients is presented in \cite{avramidi96}. The two most effective
methods that have been successfully used for the actual calculation of
the heat invariants are: 1) the functorial method of Gilkey and Branson
\cite{branson90,branson99,gilkey95}, which is based on the invariance
theory, behavior of the heat trace under conformal transformations and
some special case calculations, and 2) the method of local Taylor
expansion in normal coordinates (which is essentially equivalent  to
the geometric covariant Taylor expansions of
\cite{avramidi91b,avramidi87}). The results of both of these methods are
directly obtained in an invariant geometric form. The symbolic calculus
approach, despite being a powerful analytical tool, fails to provide
such invariant results. It gives answers in local coordinates that are
not invariant and cannot be made invariant directly. For high-order
coefficients the problem of converting such results in a geometric
invariant form is hopeless. One cannot even decide whether a particular
coefficient is zero or not. 

One of the main problems in the study of spectral asymptotics is to
develop a procedure that respects all the invariance transformations
(diffeomorphisms and gauge transformations in the physics language) of
the differential operator. Symbolic calculus gives an answer in terms of
jets of the symbol of the operator in some local coordinates. Thus there
remains a very important problem of converting these local expressions
to global geometric invariant structures, like polynomials in curvatures
and their covariant derivatives. For a general coefficient this problem
becomes  unmanagable; it is simply exponentially bad in the order of the
heat kernel coefficient. The number of the jets of the symbol is much
greater than the number of invariant structures of given order. This
problem is so bad that it is, in fact, much easier to compute the
coefficients by some other methods that directly give an invariant
answer than to use the  results of the symbolic approach. To our
knowledge, none of the results for the explicit form of the spectral
invariants were obtained by using the symbolic calculus. 

Every problem in geometric analysis has two aspects: an analytical
aspect and a geometric aspect. In the study of spectral asymptotics of
differential operators the analytic aspect has been succesfully solved
in the classical works of Greiner \cite{greiner71} and Seeley
\cite{seeley69} and others \cite{grubb96,booss93}.

The geometric aspect of the problem for Laplace type operators is now
also well understood due to the work of Gilkey \cite{gilkey75} and many
others (see
\cite{gilkey95,vassilevich03,kirsten01,berline92,avramidi02b}).
The leading symbol of a Laplace type operator naturally defines a
Riemannian metric on the manifold, which enables one to employ powerful
methods of differential geometry. In other words,  the Riemannian
structure on a manifold is determined by a Laplace type operator. We
take this fact seriously: geometry (Riemannian structure) is determined 
by analysis (differential operator). In some sense, analysis is primary
and geometry is secondary. What kind of geometry is generated does, of
course, depend on the differential operator. {\it A Laplace type
differential operator  generates the Riemannian geometry}.

As a result, much is known about the spectral asymptotics of Laplace
type operators, both on manifolds without boundary and on manifolds with
boundary, with various boundary conditions, such as Dirichlet, Neumann,
Robin, mixed, oblique, Zaremba etc. On manifolds without boundary all
odd coefficients vanish, $A_{2k+1}=0$, and all even coefficients
$A_{2k}$ up to $A_8$ have been computed in our PhD thesis
\cite{avramidi87}, which was published later as a book \cite{avramidi00}
(see also 
\cite{gilkey95,avramidi91b,avramidi99b,avramidi04c,yajima04},  the
reviews \cite{avramidi99,avramidi02b,vassilevich03} and references
therein). By using our method \cite{avramidi91b} Yajima et
al. \cite{yajima04} computed the coefficient $A_{10}$ recently.  Of
course, this remarkable progress can only be achieved by employing
modern computer algorithms (the authors of \cite{yajima04} used a
Mathematica package). The main reason for this progress is that the 
heat kernel coefficients are polynomial in the jets of the symbol of the
operator (which can be expressed in terms of curvatures and their
covariant derivatives); it is essentially an algebraic problem.

On manifolds with boundary, the heat invariants depend on the boundary
conditions. For the classical boundary conditions, like Dirichlet,
Neumann, Robin, and mixed combination thereof on vector bundles, the
coefficients $A_k$ have been explicitly computed up to $A_5$ (see, for
example, \cite{kirsten98,branson90,branson99,avramidi93}). 

A more general scheme, called oblique boundary value problem
\cite{grubb74,gilkey83a,gilkey83}, which includes tangential derivatives
along the boundary, was studied in 
\cite{avramidi98,avramidi99a,avramidi99b,dowker97,dowker99}.
This problem is not automatically elliptic like the classical boundary 
problems; there is a certain condition on the leading symbol of the
boundary operator that ensures the strong ellipticity of the problem. As
a result, the  heat invariants are no longer polynomial in the jets of
the boundary operator, which makes this problem much more difficult to
handle. So far, in the general case only the coefficient $A_1$ is known
\cite{avramidi98,avramidi99a}. In a particular
Abelian case the coefficients $A_2$ and $A_3$ have been computed
in \cite{dowker99}.

A discontinuous boundary value problem, the so-called Zaremba problem,
which includes Dirichlet boundary conditions on one part of the boundary
and Neumann boundary conditions on another part of the boundary, was
studied recently in \cite{avramidi04c,seeley01,dowker01}. Because this
problem is not smooth, the analysis becomes much more subtle (see
\cite{avramidi04c,seeley01} and references therein). In particular,
there is a singular subset of codimension $2$ on which the boundary
operator is discontinuous, and, one has to put an additional boundary
condition  that fixes the behavior at that set.   Seeley \cite{seeley01}
showed that there are no logarithmic terms in the asymptotic expansion
of the trace of the heat kernel, which are possible on general grounds,
and that the heat  invariants do depend on the boundary condition at the
singular set; the neglect of that simple fact lead to some controversy
on the coefficient $A_2$ in the past until this question was finally
settled in \cite{seeley01,avramidi04c}.

Contrary to the Laplace type operators,  {\it there are no systematic
effective methods for an explicit calculation of the heat invariants for
second-order operators which are not of Laplace type}.  Such operators
appear in so-called {\it matrix geometry}
\cite{avramidi03,avramidi04a,avramidi04b,avramidi05}, when instead
of a single Riemannian metric there is a matrix-valued symmetric
2-tensor, which we call a ``non-commutative metric''.  Matrix geometry  
is motivated by the relativistic interpretation of gauge theories and is
intimately related to Finsler geometry (rather a collection of Finsler
geometries) (see \cite{avramidi03,avramidi04a,avramidi04b}).  For an
introduction to Finsler geometry see \cite{rund59}. 

Of course, the existence and the form of the asymptotic expansion of the
heat kernel is well established for a very large class of operators,
including all self-adjoint elliptic partial differential operators with
positive definite leading symbol; it is essentially the same for all
second-order operators, whether of Laplace type or not.  However, a {\it
non-Laplace type operator does not induce a unique Riemannian metric} on
the manifold. Of course, one can pick any Riemannian metric and work
with it, but this is not natural; it does not reflect the properties of
the differential operator and its leading symbol.  Therefore, it is
useless to try to use a Riemannian structure to cast the heat invariants
in an invariant form.  Rather, a non-Laplace type operator defines a 
collection of Finsler geometries (a matrix geometry in the terminology
of \cite{avramidi03,avramidi04a,avramidi04b,avramidi05}).  Therefore, it
is the {\it matrix geometry that should be used to study the geometric
structure of the spectral invariants of non-Laplace type operators}.
This fact complicates the calculation of spectral asymptotics
significantly. Of course, the general classical algorithms described in
\cite{greiner71,seeley69} still apply.

Three decades ago Greiner \cite{greiner71}, p. 164,  indicated that {\it
``the problem of interpreting these coefficients geometrically remains
open''}. There has not been much progress in this direction. In this
sense, the study of geometric aspects of spectral asymptotics of
non-Laplace type operators is just beginning and the corresponding 
methodology is still underdeveloped in comparison with the Laplace type
theory. The only exception to this is the case of exterior  $p$-forms,
which is pretty simple and, therefore, is well understood now
\cite{gilkey91,branson94,branson97}. Thus, the {\it geometric aspect of
the spectral asymptotics of non-Laplace type operators remains an open
problem}. 

A first step in this direction was made in our papers
\cite{avramidi01,avramidi02a}. We studied a subclass of so-called 
natural non-Laplace type operators on Riemannian manifolds, which 
appear, for example, in the study of  spin-tensor quantum gauge fields.
The natural non-Laplace type operators are a special case of non-Laplace
type operators whose leading symbol is built in a universal, polynomial
way,  using tensor product and contraction from the Riemannian metric,
its inverse, together with (if applicable) the volume form and/or the
fundamental tensor-spinor.  These operators act on sections of
spin-tensor bundles. These bundles may be characterized as those
appearing as  direct summands of iterated  tensor products of the
tangent, the cotangent and the spinor bundles (see sect 2.1). 
Alternatively, they may be described abstractly as bundles associated to
representations of the spin group. These are extremely interesting and
important bundles, as they describe the fields in quantum field theory. 
The connection on the spin-tensor  bundles is built in a canonical way
from the Levi-Civita connection. The symbols of natural operators are
constructed from the jets of the Riemannian metric, the leading symbols
being constructed just from the metric. In this case, even if the
leading symbol is not scalar, its determinant is a polynomial in
$|\xi|^2=g^{\mu\nu}(x)\xi_\mu\xi_\nu$, and, therefore, its  eigenvalues
are functions of $|\xi|$ only. This allows one to use  the Riemannian
geometry and simplifies the study  of such operators significantly.

For non-Laplace operators on manifolds without boundary even the
invariant $A_4$ is not known, in general (for some partial results see
\cite{avramidi01,avramidi04b,avramidi05} and the review
\cite{avramidi02a}).  For natural non-Laplace type differential
operators on manifolds without boundary the coefficients $A_0$ and $A_2$
were computed in~\cite{avramidi01}. For general non-Laplace type
operators they were computed in \cite{avramidi04b,avramidi05}. 

The primary goal of the present work is to generalize this study to
general non-Laplace type operators on manifolds with boundary.  We
introduce a ``non-commutative'' Dirac operator as a first-order 
elliptic partial differential operator such that its square is a
second-order self-adjoint elliptic operator with positive definite
leading symbol (not necessarily of Laplace type) and study  the spectral
asymptotics of these operators with Dirichlet boundary conditions.

This paper is organized as follows. In sect. 2 we describe  the
construction of non-Laplace type operators. In sect. 2.1 we define 
natural non-Laplace type operators in the context of Stein-Weiss
operators \cite{branson97}. In sect. 2.2 we describe  a class of 
non-Laplace type operators that appear in matrix geometry following
\cite{avramidi04b,avramidi05}; we develop what can be called the
non-commutative exterior calculus and construct first-order and
second-order invariant differential operators. In sect. 2.3 we describe
the general setup of the Dirichlet boundary value problem for such
an operator and introduce necessary  tools for the analysis of the
ellipticity condition. In sect. 3 we review the spectral asymptotics of
elliptic operators both from the heat kernel and the resolvent point of
view. In sect. 4 we develop a formal technique for calculation of the
heat kernel asymptotic expansion. In sect. 4.1 the interior coefficients
$A_0$ and $A_2$ are computed (which are the same as for the manifolds
without boundary), and in the sect. 4.2 we compute the  boundary
coefficient $A_1$. 

\section{Non-Laplace Type Operators}
\setcounter{equation}0

\subsection{Natural non-Laplace type Operators}

Natural non-Laplace type operators can be constructed as follows
\cite{branson97}. Let $M$ be a smooth compact orientable $n$-dimensional
spin manifold (with or without boundary). Let $\mathcal{S}$ be the
spinor bundle over a spin manifold $M$ and
\be
\mathcal{V}=TM\otimes\cdots\otimes TM\otimes T^*M\otimes \cdots \otimes
T^*M\otimes \mathcal{S}
\ee
be a spin-tensor vector bundle corresponding to a representation of the
spin group $\mathrm{Spin}(n)$ and 
\be
\nabla: C^\infty(\mathcal{V})\to C^\infty(T^*M\otimes \mathcal{V})
\ee
be a connection on $\mathcal{V}$.  Then the decomposition 
\be
T^*M\otimes
\mathcal{V}=\mathcal{W}_1\oplus\cdots\oplus \mathcal{W}_s
\ee
of the bundle $T^*M\otimes \mathcal{V}$  into its irreducible components
$\mathcal{W}_1,\dots, \mathcal{W}_s$ defines the projections 
\be
P_j:
T^*M\otimes \mathcal{V}\to \mathcal{W}_j
\ee
and the first-order differential operators 
\be
G_j=P_j\nabla: C^\infty(\mathcal{V})\to 
C^\infty(\mathcal{W}_j),
\ee
called Stein-Weiss operators (or simply the gradients).  The number $s$
of gradients is a representation-theoretic invariant of the bundle
$\mathcal{V}$.

Then every first-order $\mathrm{Spin}(n)$-invariant differential
operator 
\be
D: C^\infty(\mathcal{V})\to  C^\infty(\mathcal{V})
\ee
is a direct sum  of the gradients 
\be
D=c_1 G_1+\cdots+c_s G_s=P\nabla\,,
\ee 
where $c_j$ are some real constants and 
\be
P=\sum_{j=1}^s c_jP_j\,,
\ee
and the second-order operators  
\be
L: C^\infty(\mathcal{V})\to  C^\infty(\mathcal{V})
\ee
defined by 
\be
L=D^*D=\nabla^* P^2\nabla=\sum_{j=1}^s c^2_j G_j^* G_j
\ee
are natural non-Laplace type operators. If all $c_j\ne 0$, then $L$ is
elliptic and has a positive definite leading symbol.

\subsection{Non-commutative Laplacian and Dirac Operator in 
Matrix Geometry}

Let $M$ be a smooth compact orientable $n$-dimensional spin manifold
with smooth boundary $\partial M$. We label the local coordinates
$x^\mu$ on the manifold $M$ by Greek indices which run over $1,\dots,
n$, and the local coordinates $\hat x^i$ on the boundary $\partial M$ by
Latin indices which run over $1,\dots, n-1$. We use the standard
coordinate bases for the tangent and the cotangent bundles. The
components of tensors over $M$ in the coordinaate basis will be labeled
by Greek indices and the components of tensors over $\partial M$ in the
coordinate basis will be labeled by Latin indices. We also use the
standard Einstein summation convention for repeated indices.

Let $\mathcal{S}$ be now an arbitrary  $N$-dimensional complex vector
bundle over $M$ (non necessarily the spinor bundle) with a positive
definite Hermitean inner product $\langle \;,\,\rangle$, $\mathcal{S}^*$
be its dual bundle and $\End(\mathcal{S})$ be the bundle of linear 
endomorphisms of the vector bundle $S$. Further, let 
$\Aut(\mathcal{S})$ be the group of automorphisms of the vector bundle
$\mathcal{S}$ and $G(\mathcal{S})$ be the group of unitary endomorphisms
of the bundle $\mathcal{S}$. We will call the unitary endomorphisms of
the  bundle $\mathcal{S}$ simply gauge transformations.

Let  $TM$ and $T^*M$ be the tangent and the contangent bundles. We
introduce the following notation for the vector bundles of vector-valued
and endomorphism-valued $p$-forms and $p$-vectors 
\be
\Lambda_p=(\wedge^p T^*M)\otimes \mathcal{S}\,,\qquad
\Lambda^p=(\wedge^p TM)\otimes \mathcal{S}\,,
\ee
\be
E_p=(\wedge^p T^*M)\otimes \End(\mathcal{S})\,,\qquad
E^p=(\wedge^p TM)\otimes \End(\mathcal{S})\,.
\ee

We will also consider vector bundles of densities of different weights 
over the manifold $M$. For each bundle we indicate the weight explicitly
in the notation of the vector bundle; for example, $\mathcal{S}[w]$ is a
vector bundle of densities of weight $w$. 

Since $M$ is orientable there is the standard volume form
$\vol=dx=dx^1\wedge \cdots\wedge dx^n$ given by the standard Lebesgue
measure in a local chart.  The volume form is, of course, a
density of weight $1$, and, hence, is a section of the bundle $E_n[1]$.
The components of the volume form in a local coordinate basis are
given by the completely anti-symmetric Levi-Civita symbol
$\varepsilon_{\mu_1\dots\mu_n}$. The $n$-vector dual to the volume form
is a density of weight $(-1)$ and, hence, is a section of the bundle
$E^n[-1]$. Its components are given by the contravariant Levi-Civita
symbol $\varepsilon^{\mu_1\dots\mu_n}$. These objects naturally define
the maps
\be
\varepsilon: \Lambda^p[w]\to \Lambda_{n-p}[w+1]\,,\qquad
\tilde\varepsilon: \Lambda_p[w]\to \Lambda^{n-p}[w-1]\,.
\ee
It is not difficult to see that
\be
\varepsilon\tilde\varepsilon=\tilde\varepsilon\varepsilon
=(-1)^{p(n-p)}{\rm Id}\,.
\ee

Further, we define the diffeomorphism-invariant $L^2$-inner product on
the space $C^\infty\left(\Lambda_p\left[\frac{1}{2}\right]\right)$ 
of smooth endomorphism-valued $p$-form densities
of weight $\frac{1}{2}$ 
by
\begin{equation}
(\psi,\varphi)=\int\limits_M dx\,\left\langle\psi(x),
\varphi(x)\right\rangle\,.
\end{equation} 
The completion of
$C^\infty\left(\Lambda_p\left[\frac{1}{2}\right]\right)$ in this norm
defines the Hilbert space
$L^2\left(\Lambda_p\left[\frac{1}{2}\right]\right)$.

Suppose we are given a map
\be
\Gamma: T^*M\to \End(\mathcal{S})
\ee
determined by a self-adjoint endomorphism-valued vector
\linebreak
$\Gamma\in C^\infty\left(TM\otimes\End(\mathcal{S})[0]\right)$,
which is described locally by a matrix-valued vector
$\Gamma^\mu$. Let us define an endomorphism-valued tensor $a\in
C^\infty\left(TM\otimes TM\otimes \End(\mathcal{S})[0]\right)$ by
\begin{equation}
a(\xi_1,\xi_2)=\frac{1}{2}\left[\Gamma(\xi_1)\Gamma(\xi_2)
+\Gamma(\xi_2)\Gamma(\xi_1)\right]\,,
\end{equation}
where $\xi_1,\xi_2\in T^*M$.
Then $a$ is self-adjoint and symmetric
\be
a(\xi_1,\xi_2)=a(\xi_2,\xi_1)\,,\qquad
\overline{a(\xi_1,\xi_2)}=a(\xi_2,\xi_1)\,.
\ee

One of our main assumptions about the tensor $a$ is that it defines an
isomorphism 
\begin{equation}
a: T^*M\otimes \mathcal{S}\to 
TM\otimes \mathcal{S}\,.
\label{16}
\end{equation}
Let us consider the endomorphism
\begin{equation}
H(x,\xi)=a(\xi,\xi)=[\Gamma(\xi)]^2\,,
\label{24}
\end{equation}
with $x\in M$, and $\xi\in T_x^*M$ being a cotangent vector. 
Our second assumption is that this endomorphism
is positive definite, i.e.
\be
H(x,\xi)>0
\ee
for any point $x$ of the manifold $M$ and $\xi\ne 0$.
This endomorphism is self-adjoint and, therefore,  all its eigenvalues
are real and positive for $\xi\ne 0$. We call the endomorphism-valued
tensor $a$ the {\it non-commutative metric} and the components
$\Gamma^\mu$ of the endomorphism-valued vector $\Gamma$ the {\it
non-commutative Dirac matrices}.

This construction determines a collection of Finsler
geometries~\cite{avramidi04a,avramidi05}.  Assume, for simplicity, that
the matrix  $H(x,\xi)=a(\xi,\xi)$ has distinct eigenvalues: 
$h_{(a)}(x,\xi)$, $a=1,\dots,N$.
Each eigenvalue defines a Hamilton-Jacobi equation
\be
h_{(a)}(x,\partial S)=m_{(a)}^2\,,
\ee
where $m_{(a)}$ are some constants,
a Hamiltonian system
\bea
\frac{dx^\mu}{dt}&=&\frac{1}{2}
\frac{\partial}{\partial \xi_\mu}h_{(a)}(x,\xi)\,,
\\
\frac{d\xi^\mu}{dt}&=&-\frac{1}{2}
\frac{\partial}{\partial x^\mu}h_{(a)}(x,\xi)\,,
\eea
(the coefficient $1/2$ is introduced here for convenience)
and a positive definite Finsler metric
\be
g_{(a)}^{\mu\nu}(x,\xi)=\frac{1}{2}
{\partial^2 h_{(a)}\over
\partial\xi_\mu\partial\xi_\nu}\,.
\ee

Moreover, each eigenvalue is a positive homogeneous function of
$\xi$ of degree $2$ and, therefore, the Finsler metric 
is a homogeneous function of $\xi$ of degree $0$.
This leads to a number of identities, in particular,
\be
h_{(a)}(x,\xi)=g^{\mu\nu}_{(a)}(x,\xi)\xi_\mu\xi_\nu
\ee
and
\be
\dot x^\mu=g^{\mu\nu}_{(a)}(x,\xi)\xi_\nu\,.
\ee
Next, one defines the inverse (covariant) Finsler metrics
\be
g_{(a)\,\mu\nu}(x,\dot x)g^{\nu\alpha}_{(a)}(x,\xi)=\delta^\alpha_\mu\,,
\ee
the interval
\be
ds_{(a)}^2=g_{(a)\,\mu\nu}(x,\dot x)\, dx^\mu dx^\nu\,,
\ee
connections, curvatures etc (for details,
see~\cite{rund59}).
Thus, a {\it non-Laplace type operator generates a collection of 
Finsler geometries}.

The isomorphism $a$ naturally defines a map
\begin{equation}
A: \Lambda_p\to \Lambda^{p}\,,
\end{equation}
by
\begin{equation}
(A\varphi)^{\mu_1\cdots\mu_p} =A^{\mu_1\cdots\mu_p\nu_1\cdots\nu_p}
\varphi_{\nu_1\cdots\nu_p}\,,
\end{equation}
where
\begin{equation}
A^{\mu_1\cdots\mu_p\nu_1\cdots\nu_p}=
\delta_{\alpha_1}^{[\mu_1}\cdots\delta_{\alpha_p}^{\mu_p]}
\delta_{\beta_1}^{[\nu_1}\cdots\delta_{\beta_p}^{\nu_p]}
a^{\alpha_1\beta_1}\cdots
a^{\alpha_p\beta_p}\,,
\end{equation}
and the square brackets denote the complete antisymmetrization
over the indices included.
We will assume that these maps are isomorphisms as well. 
Then the inverse operator
\begin{equation}
A^{-1}: \Lambda^p\to \Lambda_{p}\,,
\end{equation}
is defined by
\begin{equation}
(A^{-1}\varphi)_{\mu_1\cdots\mu_p}
=(A^{-1})_{\mu_1\cdots\mu_p\nu_1\cdots\nu_p}
\varphi^{\nu_1\cdots\nu_p}\,,
\end{equation}
where $A^{-1}$ is determined by the equation
\begin{equation}
(A^{-1})_{\mu_1\cdots\mu_p\nu_1\cdots\nu_p}
A^{\nu_1\cdots\nu_p\alpha_1\cdots\alpha_p}
=\delta^{\alpha_1}_{[\mu_1}\cdots\delta^{\alpha_p}_{\mu_p]}\,.
\end{equation}

This can be used further to define the natural inner product on the
space of $p$-forms $\Lambda_p$ via
\begin{equation}
\left\langle\psi,\varphi\right\rangle ={1\over
  p!}\bar\psi_{\mu_1\cdots\mu_p} A^{\mu_1\cdots\mu_p\nu_1\cdots\nu_p}
\varphi_{\nu_1\cdots\nu_p}\,.
\end{equation}

Let $d$ be the exterior derivative on $p$-form densities of weight $0$
\begin{equation}
d: C^\infty(\Lambda_p[0])\to C^\infty(\Lambda_{p+1}[0])\,
\end{equation}
and $\tilde d$ be the coderivative on $p$-vector densities 
of weight $1$
\begin{equation}
\tilde d=(-1)^{np+1}\tilde\varepsilon d\varepsilon: \  
C^\infty(\Lambda^p[1])\to C^\infty(\Lambda^{p-1}[1])\,.
\end{equation}
These operators are invariant differential operators defined 
without a Riemannian metric. They take the following form
in local coordinates
\begin{eqnarray}
(d \varphi)_{\mu_1\cdots\mu_{p+1}}
&=&(p+1)\partial_{[\mu_1}\varphi_{\mu_2\cdots\mu_p]}\,,
\\[10pt]
(\tilde d \varphi)^{\mu_1\cdots\mu_{p-1}}
&=&\partial_\mu \varphi^{\mu\mu_1\cdots\mu_{p-1}}\,.
\end{eqnarray}

Now, let $\mathcal{B}\in C^\infty\left(T^*M\otimes \End(\mathcal{
S})[0]\right)$ be a smooth anti-self-adjoint endo\-mor\-phism-valued
connection $1$-form on the bundle $\mathcal{S}$,  defined by the
matrix-valued covector $\mathcal{B}_\mu$.
Such a section naturally defines the maps:
\be
\mathcal{B}: \Lambda_p\left[\textstyle\frac{1}{2}\right]\to 
\Lambda_{p+1}\left[\textstyle\frac{1}{2}\right]
\ee
and
\be
\tilde{\mathcal{B}}=(-1)^{np+1}\tilde\varepsilon \mathcal{B}\varepsilon
: \Lambda^p\left[\textstyle\frac{1}{2}\right]\to
\Lambda^{p-1}\left[\textstyle\frac{1}{2}\right]
\ee
given locally by
\bea
(\mathcal{B}\varphi)_{\mu_1\cdots\mu_{p+1}}
&=&(p+1)\mathcal{B}_{[\mu_1}
  \varphi_{\mu_2\cdots\mu_{p+1}]}\,,
\\[10pt]
(\tilde{\mathcal{B}}\varphi)^{\mu_1\cdots\mu_{p-1}} 
&=&\mathcal{B}_\mu\varphi^{\mu\mu_1\cdots\mu_{p-1}}\,.
\eea

Finally, we introduce a self-adjoint non-degenerate endomorphism-valued
density   $\rho\in
C^\infty\left(\End(\mathcal{S})\left[\frac{1}{2}\right]\right)$ of
weight $\frac{1}{2}$. Then $\rho^2$ has weight $1$ and plays the role of
a {\it non-commutative measure}.

This enables us to define the covariant exterior derivative
of $p$-form densities of weight $\frac{1}{2}$
\begin{equation}
\mathcal{D}: C^\infty\left(\Lambda_p \left[{\textstyle{1\over
      2}}\right]\right)\to C^\infty\left(\Lambda_{p+1}
\left[{\textstyle\frac{1}{2}}\right]\right).
\end{equation}
and the covariant coderivative of $p$-vector densities of 
weight $\frac{1}{2}$
\begin{equation}
\tilde{\mathcal{D}}=
(-1)^{np+1}\tilde\varepsilon \mathcal{D}\varepsilon: \ 
C^\infty\left(\Lambda^p \left[{\textstyle{1\over
      2}}\right]\right)\to C^\infty\left(\Lambda^{p-1}
\left[{\textstyle\frac{1}{2}}\right]\right),
\end{equation} 
by
\begin{equation}
\mathcal{D}=\rho(d+\mathcal{B})\rho^{-1}\,,
\end{equation}
\begin{equation}
\tilde{\mathcal{D}}=\rho^{-1}(\tilde d+\tilde{\mathcal{B}})\rho \,.
\end{equation}
These operators transform covariantly under both the diffeomorphisms
and the gauge transformations.

The formal adjoint of the operator $\mathcal{D}$ 
\begin{equation}
\bar{\mathcal{D}}: C^\infty\left(\Lambda_p \left[{\textstyle{1\over
      2}}\right]\right)\to C^\infty\left(\Lambda_{p-1}
\left[{\textstyle\frac{1}{2}}\right]\right),
\end{equation}
has the form
\begin{equation}
\bar{\mathcal{D}}
=-A^{-1}\rho^{-1}(\tilde d+\tilde{\mathcal{B}})\rho A\,,
\end{equation}
By making use of these operators we define a second-order operator 
(that can be called the {\it non-commutative Laplacian})
\begin{equation}
\Delta:\
C^\infty\left(\Lambda_p \left[{\textstyle\frac{1}{2}}\right]\right)\to
C^\infty\left(\Lambda_{p} \left[{\textstyle\frac{1}{2}}\right]\right),
\end{equation}
by
\be
\Delta=-\bar{\mathcal{D}}\,\mathcal{D}
-\mathcal{D}\,\bar{\mathcal{D}}\,.
\ee

In the special case $p=0$ the non-commutative Laplacian $\Delta$
reads
\begin{equation}
\Delta=\rho^{-1}(\tilde d+\tilde{\mathcal{B}})\rho A\rho(d+\mathcal{
  B})\rho^{-1}\,,
\end{equation}
which in local coordinates has the form
\begin{equation}
\Delta=\rho^{-1}(\partial_\mu+\mathcal{B}_\mu)\rho
a^{\mu\nu}\rho(\partial_\nu+\mathcal{B}_\nu)\rho^{-1}\,.
\end{equation}

Next, notice that the endomorphism-valued vector 
$\Gamma$ introduced above
naturally defines the maps
\be
\Gamma:\ C^\infty\left(\Lambda^p\left[{\textstyle\frac{1}{2}}\right]\right)
\to C^\infty\left(\Lambda^{p+1}\left[{\textstyle\frac{1}{2}}\right]\right)
\ee
and
\be
\tilde\Gamma=(-1)^{np+1}\varepsilon\Gamma\tilde\varepsilon:\ 
C^\infty\left(\Lambda_p\left[{\textstyle\frac{1}{2}}\right]\right)
\to C^\infty\left(\Lambda_{p-1}\left[{\textstyle\frac{1}{2}}\right]\right)
\ee
given locally by
\bea
(\Gamma\varphi)^{\mu_1\dots\mu_{p+1}}
&=&(p+1)\Gamma^{[\mu_1}\varphi^{\mu_2\dots\mu_{p+1}]}
\\[10pt]
(\tilde\Gamma\varphi)_{\mu_1\dots\mu_{p-1}}
&=&\Gamma^\mu\varphi_{\mu\mu_1\dots\mu_{p-1}}\,.
\eea

Therefore, we can define a first-order invariant differential operator
(that can be called the {\it non-commutative Dirac operator})
\begin{equation}
D: C^\infty\left(\Lambda_p \left[{\textstyle\frac{1}{2}}\right]\right)\to
C^\infty\left(\Lambda_{p} \left[{\textstyle\frac{1}{2}}\right]\right)\
\end{equation}
by
\begin{equation}
D=i\tilde\Gamma \mathcal{D} 
= i\tilde\Gamma\rho(d+\mathcal{B})\rho^{-1}\,,
\end{equation}
where, of course, $i=\sqrt{-1}$.
The formal adjoint of this operator is
\begin{equation}
\bar{D}=iA^{-1}\tilde{\mathcal{D}}\Gamma A 
= iA^{-1}\rho^{-1}(\tilde
d+\tilde{\mathcal{B}})\rho\Gamma A\,.
\end{equation}
These operators can be used to define second order differential
operators $D\bar D$ and $\bar D D$.

In the case $p=0$ these operators have the following form in local 
coordinates
\be
D=i\Gamma^\mu
\rho(\partial_{\mu}+\mathcal{B}_{\mu})\rho^{-1}\,,
\ee
\be
\bar{D} = 
i\rho^{-1}(\partial_{\nu}+\mathcal{B}_{\nu})\rho\Gamma^{\nu}\,,
\ee
and, therefore, the second-order operators $D\,\bar{D}$ and 
$\bar{D}\,D$ read
\begin{eqnarray}
D\,\bar{D}&=&-\Gamma^\mu \rho(\partial_\mu+\mathcal{B}_\mu)
\rho^{-2}(\partial_\nu+\mathcal{B}_\nu)\rho\Gamma^\nu\,,
\\[10pt]
\bar{D}\,D&=&-\rho^{-1}(\partial_\nu+\mathcal{B}_\nu)
\rho\Gamma^\nu \Gamma^\mu
\rho(\partial_\mu+\mathcal{B}_\mu)\rho^{-1}\,.
\end{eqnarray}

In the present paper we will primarily study the  second-order operators
$\Delta$, $\bar{D}\,D$ and $D\,\bar{D}$ in the case $p=0$, that is,
$$
\Delta, \bar{D}\,D, D\,\bar{D}: 
C^\infty\left(\mathcal{S}\left[{\textstyle\frac{1}{2}}\right]\right)
\to C^\infty\left(\mathcal{S}\left[{\textstyle\frac{1}{2}}\right]\right)\,.
$$  
These are all formally self-adjoint operators by construction. This
means that they are symmetric on  smooth sections of the bundle $\mathcal{
S}\left[{\textstyle\frac{1}{2}}\right]$ with compact support in the
interior of $M$ (that is, sections that vanish together with all their
derivatives on the boundary $\partial M$).

The leading symbols of all these operators are equal to the matrix
$H(x,\xi)=a(\xi,\xi)$, i.e.
\be
\sigma_L(\Delta;x,\xi)
=\sigma_L(\,\bar{D}\,D;x,\xi)
=\sigma_L(D\,\bar{D};x,\xi)
=H(x,\xi)
=a(\xi,\xi)\,,
\ee
where $\xi\in T_x^*M$. By our main assumption about the non-commuting
metric the leading symbol is self-adjoint and positive definite in the
interior of the manifold. Therefore, the leading symbol is invertible
(or elliptic) in the interior of $M$. Notice that the leading symbol is
non-scalar, in general. That is why such operators are called 
{\it non-Laplace type operators}.

\subsection{Elliptic Boundary Value Problem}

Let us consider a neighborhood of the boundary $\partial M$
in $M$. Let $x=(x^\mu)$ be the local coordinates in this neighborhood.
The boundary is  a smooth hypersurface without boundary. Therefore,
there must exist a local diffeomorphism
\be
r=r(x)\,\qquad
\hat x^i=\hat x^i(x)\,,\qquad i=1,\dots,n-1\,,
\ee
and the inverse diffeomorphism
\be
x^\mu=x^\mu(r,\hat x)\,,\qquad \mu=1,\dots,n\,,
\ee
such that
\be
r(x)=0 \qquad \mbox{for any } x\in \partial M\,,
\ee
\be
r(x)>0 \qquad \mbox{for any } x \not\in \partial M\,,
\ee
and the vector $\partial_r={\partial/\partial r}$ is {\it transversal}
(nowhere tangent) to the boundary $\partial M$. Then the coordinates
$\hat x^i$ are local coordinates on the boundary $\partial M$.

Let $\delta>0$. We define a disjoint decomposition of the manifold
\be
M=M_{\rm int}\cup M_{\rm bnd}\,,
\ee
where 
\be
M_{\rm bnd}=\{x\in M\;|\; r(x)<\delta\}\,
\ee
is a $\delta$-neighborhood of the boundary
and
\be
M_{\rm int}=M\setminus M_{\rm bnd}\,
\ee
is the part of the interior of the manifold on a finite distance from the
boundary.

For $r=0$, that is, $x\in\partial M$, the vectors $\{\hat
\partial_i={\partial/\partial \hat x^i}\}$ are tangent to the boundary and
give the local coordinate basis for the tangent space $T_x\partial M$.
The set of vectors $\{\partial_r,\hat\partial_i\}$ gives the local
coordinate basis for the tangent space  $T_xM$ in $M_{\rm bnd}$.
Similarly, the $1$-forms $d\hat x^i$ determine the local coordinate
basis for the cotangent space $T_x^*\partial M$, and the $1$-forms
$\{dr,d\hat x^i\}$ give the local coordinate basis for the cotangent
space $T^*_xM$ in $M_{\rm bnd}$.

We fix the orientation of the boundary by requiring the Jacobian  of
this diffeomorphism to be positive, in other words, for any $x\in M_{\rm bnd}$
\be
J(x)=\vol(\partial_r,\hat\partial_1,\dots,\hat\partial_{n-1})>0\,.
\ee

Let $\varphi\in C^\infty(TM[1])$ 
be a smooth vector density of weight $1$. Then
Stokes' Theorem has the form
\be
\int\limits_M dx\; \tilde d\varphi
=\int\limits_{\partial M} d\hat x\; N(\varphi)\,, 
\ee
where $N$ is a $1$-form defined by
\bea
N(\varphi)
&=&
\vol(\varphi,\hat\partial_1,\dots,\hat\partial_{n-1})
=\frac{1}{J} dr(\varphi)
\nonumber\\
&=&
\varepsilon_{\mu\nu_1\dots\nu_{n-1}}
{\partial x^{\nu_1}\over\partial \hat x^{1}}
\cdots
{\partial x^{\nu_{n-1}}\over\partial \hat x^{n-1}}
\varphi^\mu
=\frac{1}{J}{\partial r\over \partial x^\mu}\varphi^\mu\,.
\eea
Notice that this formula is valid for densities, and there is no need for 
a Riemannian metric.

We will study in the present paper, for simplicity, the 
Dirichlet boundary conditions
\be
\varphi\big|_{\partial M}=0\,.
\ee
By integration by parts it is not difficult to see that all operators
$\Delta$, $D\,\bar{D}$ and $\bar{D}\,D$ are symmetric  on
smooth sections of the bundle $\mathcal{S}\left[\frac{1}{2}\right]$
satisfying the boundary conditions. One can show that these operators
are essentially self-adjoint, that is, their closure is self-adjoint
and, hence, they have unique self-adjoint extensions to
$L^2\left(\mathcal{S}\left[{1\over 2}\right]\right)$. 

Let $L$ be one of the operators  $\bar{D}\,D, D\,\bar{D}\,,
\Delta$ with the Dirichlet boundary conditions. Our primary interest in
this paper is the study of elliptic boundary value problems. Ellipticity
means invertibility up to a compact operator in appropriate functional
spaces (see, for example, \cite{booss93,gilkey83,gilkey95}). This is,
roughly speaking, a condition that implies local invertibility. For a
boundary value problem it has two components: i) in the interior of the
manifold, and ii) at the boundary.

An operator $L$ is elliptic in the interior of the manifold if for any
interior point $x\in M$ and for any nonzero cotangent vector $\xi\in
T^*_xM$, $\xi\ne 0$, its leading symbol $\sigma_L(L;x,\xi)$ is
invertible. Since all operators $D\,\bar{D}$, $\bar{D}\, D$
and $\Delta$ all have positive leading symbols, namely $H(x,\xi)$, they
are elliptic in the interior of the manifold.

At the boundary $\partial M$ of the manifold we use the coordinates
$(r,\hat x)$ and define a split of the cotangent bundle $T^*M=\RR\oplus
T^*\partial M$, so that $\xi=(\xi_\mu)=(\omega,\hat\xi)\in T^*_xM$,
where $\omega \in\RR$ and $\hat\xi=(\xi_j)\in T^*_x\partial M$.  

Let $\lambda\in \CC\setminus\RR_+$ be a complex number that does not lie
on the positive real axis and $H(r,\hat x, \omega,\hat \xi)$ be the
leading symbol of the operator $L$. We substitute $r=0$ and
$\omega\mapsto-i\partial_r$ and consider the following  second-order
ordinary differential equation on the half-line, i.e. $r\in\RR_+$,
\be
\left[H(0,\hat x,-i\partial_r,\hat\xi)-\lambda\right]\varphi=0\,,
\label{bode}
\ee
with an asymptotic condition
\be
\lim_{r\to \infty}\varphi=0\,.
\label{basym}
\ee

Let $\hat{\mathcal{S}}=\mathcal{S}|_{\partial M}$ be the restriction of
the vector bundle $\mathcal{S}$ to the boundary.
The operator $L$ with Dirichlet boundary conditions is elliptic
if  for each boundary point $\hat x\in \partial M$, each $\hat\xi\in
T^*_{\hat x}\partial M$, each $\lambda\in \CC\setminus \RR_+$, such that
$\hat
\xi$ and $\lambda$ are not both zero, and each $f\in
C^\infty\left(\hat{\mathcal{S}}\left[\frac{1}{2}\right]\right)$ there is
a unique solution $\varphi(\lambda,r,\hat\xi)$ to the equation
(\ref{bode}) subject to the asymptotic condition (\ref{basym}) at
infinity and the boundary condition at $r=0$
\be
\varphi(\lambda,0,\hat \xi)=f\,.
\label{bcon}
\ee

We have
\bea
H(0,\hat x,\omega,\hat\xi)&=&[A(\hat x)\omega+C(\hat x,\hat\xi)]^2
\nonumber\\
&=&A^2(\hat x)\omega^2+B(\hat x,\hat \xi)\omega
+C^2(\hat x,\hat \xi)\,,
\eea
where $A$, $B$, and $C$ are self-adjoint matrices defined by
\be
A(\hat x)=\Gamma(dr)\,, \qquad 
C(\hat x,\hat\xi)=\Gamma(d\hat x^j)\hat\xi_j\,,
\label{matac}
\ee
\be
B(\hat x,\hat\xi)=A(\hat x)C(\hat x,\hat \xi)
+C(\hat x,\hat\xi)A(\hat x)\,.
\label{matb}
\ee
Then the differential equation (\ref{bode}) has the form
\be
\left(-A^2\partial_r^2-iB\partial_r+C^2-\lambda\II\right)\varphi=0\,.
\label{bode1}
\ee

We notice that the matrix $\left[(A\omega+C)^2-\lambda\II\right]$
is non-degenerate when $\omega$ is real and $\lambda$ and $\hat\xi$ are
not both zero, i.e.  $(\lambda,\hat\xi)\ne(0,0)$. Moreover, when
$\lambda$ is  a negative real number, then this matrix is self-adjoint
and positive definite for real $\omega$. 
Therefore, we can define
\be
\Phi(\lambda,y,\hat\xi)
=\int\limits_{-\infty}^\infty \frac{d\omega}{2\pi}e^{i\omega y}
R_\lambda(\omega,\hat\xi)\,,
\label{kdef}
\ee
where
\be
R_\lambda(\omega,\hat\xi)
=\left\{[A(\hat x)\omega+C(\hat x,\hat\xi)]^2-\lambda\II\right\}^{-1} \,.
\ee
The matrix $\Phi(\lambda,y,\hat\xi)$ is well defined for 
any $y\in\RR$. It: i) vanishes at infinity,
\be
\lim_{y\to\pm\infty}\Phi(\lambda,y,\hat\xi)=0\,,
\ee
ii) satisfies the symmetry relations
\be
\overline{\Phi(\lambda,y,\hat\xi)}
=\Phi(\bar\lambda,-y,\hat\xi)\,,
\qquad
\Phi(\lambda,y,-\hat\xi)
=\Phi(\lambda,-y,\hat\xi)\,,
\label{phixi}
\ee
iii) is homogeneous, i.e. for any $t>0$,
\be
\Phi\left(\frac{\lambda}{t},\sqrt{t}\,y,\frac{\hat\xi}{\sqrt{t}}\right)
=t^{1/2}\Phi(\lambda,y,\hat\xi)\,,
\ee
iv) is continuous at zero with a well defined value at $y=0$
\be
\Phi_0(\lambda,\hat\xi)=\Phi(\lambda,0,\hat\xi)
=\int\limits_{-\infty}^\infty \frac{d\omega}{2\pi}
R_\lambda(\omega,\hat\xi)\,,
\label{73}
\ee
v) has a discontinuous derivative $\partial_y\Phi(\lambda,y,\hat\xi)$
at $y=0$ with a finite jump.

We also notice that the matrix $\Phi_0(\lambda,\hat\xi)$ is an 
even function of $\hat\xi$ and 
is self-adjoint for real $\lambda$, i.e.
\be
\Phi_0(\lambda,-\hat\xi)=\Phi_0(\lambda,\hat\xi)\,,\qquad
\overline{\Phi_0(\lambda,\hat\xi)}
=\Phi_0(\bar\lambda,\hat\xi)\,.
\ee
Moreover, for real negative $\lambda$ the matrix $\Phi_0$ is positive and,
therefore, non-degenerate. More generally, it is non-degenerate for
$\mathrm{Re}\,\lambda<w$, where $w$ is a sufficiently large negative
constant.

In an important particular case, when $B=AC+CA=0$, one can compute explicitly
\be
\Phi(\lambda,y,\hat\xi)
=\frac{1}{2}A^{-1}\mu^{-1}e^{-\mu |y|}A^{-1}\,,
\qquad
\Phi_0(\lambda,\hat\xi)
=\frac{1}{2}A^{-1}\mu^{-1}A^{-1}\,,
\ee
where $\mu=\sqrt{A^{-1}(C^2-\lambda\II)A^{-1}}$, defined as an 
analytical continuation in $\lambda$ of a positive square 
root of a self-adjoint matrix when $\lambda\in \RR_-$.

One can prove now that the eq. (\ref{bode1}) with initial condition
(\ref{bcon}) and the asymptotic condition at infinity (\ref{basym})
has a unique solution given by
\be
\varphi(\lambda,r,\hat\xi)
=\Phi(\lambda,r,\hat\xi)[\Phi_0(\lambda,\hat\xi)]^{-1}
f\,.
\ee
Thus, the Dirichlet boundary value problem for our operator is elliptic.

\section{Spectral Asymptotics}
\setcounter{equation}0
\subsection{Heat Kernel}

Let $L$ be a self-adjoint elliptic second-order partial differential
operator acting on smooth sections of the bundle $\mathcal{S}\left[{1\over
2}\right]$ over a compact manifold $M$ with boundary $\partial M$ with
positive definite leading symbol and with some boundary conditions
\be
B\varphi|_{\partial M}=0\,,
\ee
with some boundary operator $B$.
It is well known that such an operator
has a discrete real spectrum $\{\lambda_k\}_{k=1}^\infty$ bounded from
below~\cite{gilkey95}, i.e.,
\be
\lambda_1\le\lambda_2\le\cdots\le \lambda_{k}\le\lambda_{k+1}\le\cdots\,.
\ee 
Furthermore: i) each eigenspace is finite-dimensional, ii) the
eigensections are smooth sections of the bundle  $\mathcal{S}\left[{1\over
2}\right]$, and iii) the set of eigensections
$\{\varphi_k\}_{k=1}^\infty$ forms an orthonormal basis in 
$L^2\left(\mathcal{S}\left[\frac{1}{2}\right]\right)$.
 
For $t>0$ the heat semigroup 
\be
\exp(-tL): L^2\left(\mathcal{S}\left[\textstyle\frac{1}{2}\right]\right)\to 
L^2\left(\mathcal{S}\left[\textstyle\frac{1}{2}\right]\right)
\ee
is a bounded operator. 
The integral kernel of this operator, called the heat kernel, is
given by
\be
U(t;x,x')=\sum_{k=1}^\infty e^{-t\lambda_k}\varphi_k
\otimes\bar\varphi_k(x')\,,
\ee
where each eigenvalue is counted with its multiplicity.
The heat kernel satisfies the heat equation
\be
(\partial_t+L)U(t;x,x')=0
\ee
with the initial condition
\be
U(0^+;x,x')=\delta(x,x')\,,
\ee
where $\delta(x,x')$ is the Dirac distribution, as well as the
boundary conditions
\be
B_xU(t;x,x')\Big|_{x\in \partial M}=0\,,
\ee
and the self-adjointness condition
\be
\overline{U(t;x,x')}=U(t;x',x)\,.
\ee

The heat kernel $U(t)=\exp(-tL)$ is intimately related to the resolvent
$G(\lambda)=(L-\lambda)^{-1}$. Let $\lambda$ be
a complex number with a sufficiently large negative real part, 
$\mathrm{Re}\lambda<<0$. 
Then the resolvent and the heat kernel are related by the 
Laplace transform
\be
G(\lambda)=\int\limits_0^\infty dt\; e^{t\lambda}\, U(t)\,,
\ee
\be
U(t)=\frac{1}{2\pi i}\int\limits_{w-i\infty}^{w+i\infty} d\lambda\;
e^{-t\lambda}\,G(\lambda)\,,
\ee
where $w$ is a sufficiently large negative real number, $w<<0$.

The resolvent satisfies the equation
\be
(L-\lambda\II)G(\lambda;x,x')=\delta(x,x')
\ee
with the boundary condition
\be
B_x G(\lambda;x,x')\big|_{x\in\partial M}=0\,,
\ee
and the self-adjointness condition
\be
\overline{G(\lambda;x,x')}=G(\bar\lambda;x',x)\,.
\ee
The integral kernel of the resolvent reads
\be
G(\lambda;x,x')=\sum_{k=1}^\infty \frac{1}{\lambda_k-\lambda}\;\varphi_k
\otimes\bar\varphi_k(x')\,,
\ee
where each eigenvalue is counted with its multiplicity.

For $t>0$ the heat kernel $U(t;x,x')$ is a smooth  section of the
exterior tensor product bundle  $\mathcal{S}\left[{1\over
2}\right]\boxtimes \mathcal{S}^*\left[\frac{1}{2}\right]$; that is, it is a
two-point density of weight $\frac{1}{2}$ at each point. In particular,
it is a smooth function near the diagonal of $M\times M$ and has a well
defined diagonal value $U(t;x,x)$. The diagonal is, of course, a smooth
section of the bundle $\mathcal{S}\left[1\right]$, a density of weight $1$.

Moreover, the heat semigroup is a trace-class operator with a well
defined $L^2$-trace
\be
\Tr_{L^2}\exp(-tL)=\int\limits_M dx\; \tr_{S}U(t;x,x)\,,
\ee
where $\tr_{S}$ is the trace over the fiber vector space $S$
of the vector bundle $\mathcal{S}$.
The trace of the heat kernel is a spectral invariant of the operator
$L$ since
\be
\Tr_{L^2}\exp(-tL)=\sum_{k=1}^\infty e^{-t\lambda_k}\,.
\ee
Since the diagonal is a density of weight $1$ the trace 
$\Tr_{L^2}\exp(-tL)$ is invariant under diffeomorphisms. 

This enables one to define other spectral functions by integral
transforms of the trace of the heat kernel. In particular, the  zeta
function, $\zeta(L;s,\lambda)$, is defined as follows. Let $\lambda$ be
a complex parameter with ${\rm Re}\,\lambda<\lambda_1$, so that the operator
$(L-\lambda)$ is positive. Then for any $s\in \CC$ such that
${\rm Re}\,s>n/2$ the trace of the operator $(L-\lambda)^{-s}$ is well
defined and determines
the zeta function,
\be
\zeta(L;s,\lambda)=\Tr_{L^2}(L-\lambda)^{-s}
={1\over\Gamma(s)}\int\limits_0^\infty dt\; t^{s-1}\,
e^{t\lambda}\,\Tr_{L^2}\exp(-tL)\,.
\ee
The zeta function enables one to define further the regularized
determinant of the operator $(L-\lambda)$ by
\be
{\partial\over\partial s}
\zeta(L;s,\lambda)\Big|_{s=0}
=-\log\Det (L-\lambda)\,.
\ee

There is an asymptotic expansion
as $t\to 0$ of the trace of the
heat kernel \cite{gilkey95} (for a review, see also 
\cite{avramidi91b,avramidi99,avramidi00,avramidi02b,vassilevich03})
\be
\Tr_{L^2}\exp(-tL)
\sim (4\pi)^{-n/2}\sum_{k=0}^\infty t^{(k-n)/2} A_k(L)\,.
\ee
The coefficients $A_{k}(L)$, called the global heat invariants,
are spectral invariants determined by the integrals over the manifold 
$M$ and the boundary $\partial M$ of some scalar densities
$a_k(L;x)$ and $b_k(L;\hat x)$, called local heat invariants,
viz.
\be
A_k(L)=\int\limits_M dx\; a_k(L;x)
+\int\limits_{\partial M} d\hat x\;b_k(L;\hat x)\,.
\ee
The local heat invariants $a_k(L;x)$ and $b_k(L;\hat x)$ 
are constructed polynomially from the jets of the
symbol  of the operator $L$; the boundary coefficients $b_k$
depend, of course, on the boundary conditions and the geometry
of the boundary as well.

Contrary to the heat kernel, the resolvent is singular at the
diagonal and does not have a well defined trace. However, the
derivatives of the resolvent do. Let $m\ge n/2$. Then the trace
$\Tr_{L^2}(\partial_\lambda)^m G(\lambda)$ is well defined and has
the asymptotic expansion as $\lambda\to -\infty$
\be
\Tr_{L^2}\frac{\partial^m}{\partial\lambda^m}G(\lambda)
\sim (4\pi)^{-n/2}\sum_{k=0}^\infty 
\Gamma\left[(k-n+2m+2)/2\right]
(-\lambda)^{(n-k-2m-2)/2}A_k(L)\,.
\ee
Therefore, one can use either the heat kernel or the resolvent to compute
the coefficients $A_k$.

\subsection{Index of Noncommutative Dirac Operator}

Notice that the operator $\Delta$ can have a finite number of negative
eigenvalues, whereas the spectrum of the operators $\bar{D}\,D$ and 
$D\,\bar{D}$ is non-negative. 
Moreover, one can easily show that all non-zero
eigenvalues of the operators $\bar{D}\,D$ and $D\,\bar{D}$ are equal
\be
\lambda_k(\,\bar{D}\,D)=\lambda_k(D\,\bar{D}) \qquad 
\mbox{if } \lambda_k(\,\bar{D}\,D)>0\,.
\ee
Therefore, there is a well defined index
\be
{\rm Ind}(D)=\dim\Ker(\,\bar{D}\,D)-\dim\Ker(D\,\bar{D})\,,
\ee
which is equal to the difference of the number of 
zero modes of the operators $\bar{D}\,D$
and $D\,\bar{D}$.

This leads to the fact that the difference of the heat traces
for the operators $\bar{D}\,D$ and $D\,\bar{D}$
determines the index
\be
\Tr_{L^2}\exp(-t\,\bar{D}\,D)
-\Tr_{L^2}\exp(-t\,D\,\bar{D}\,)
=\Ind(D)\,.
\ee
This means that the spectral invariants of the
operators $\bar{D}\,D$ and $D\,\bar{D}$ are equal except for the
invariant $A_{n}$ which determines the index
\be
A_k(\,\bar{D}\,D)=A_k(D\,\bar{D}\,) 
\qquad \mbox{for } k\ne n\,,
\ee
and
\be
A_n(\,\bar{D}\,D)-A_n(D\,\bar{D}\,)
=(4\pi)^{n/2}\Ind(D)\,.
\ee

Thus, for $n>2$ the spectral invariants $A_0$, $A_1$ and $A_2$ of the
operators $\bar{D}\,D$ and  $D\,\bar{D}$ are equal. Therefore,
we can pick any of these operators to compute these invariants. Of
course, the spectral invariants of the noncommutaative Laplacian
$\Delta$ are, in general, different. However, since the operators 
$\bar{D}\,D$ and  $D\,\bar{D}$ have the same leading symbol as
the operator $\Delta$ there must exist a corresponding
Lichnerowicz-Weitzenbock formula (for the spinor bundle see, for
example, \cite{berline92}), which means that the spectral invariants of
these operators must be related.

\section{Heat Invariants}
\setcounter{equation}0
\subsection{Interior Coefficients}

The heat kernel in the interior part is constructed as follows.
We fix a point $x_0\in M_{\rm int}$ in the interior of the manifold and
consider a neighborhood of $x_0$ disjoint from the boundary layer
$M_{\rm bnd}$ covered by a single patch of local coordinates.
We introduce a scaling parameter $\varepsilon>0$ and 
scale the variables according to
\be
x^\mu\mapsto x^\mu_0+\varepsilon(x^\mu-x^\mu_0)\,,\qquad 
x'^\mu\mapsto x^\mu_0+\varepsilon(x'^\mu-x^\mu_0)\,,\qquad
t\mapsto\varepsilon^2 t\,,
\ee
so that
\be
\partial_\mu \mapsto\frac{1}{\varepsilon}\partial_\mu\,,\qquad
\partial_t \mapsto\frac{1}{\varepsilon^2}\partial_t\,.
\ee

Then the differential operator 
$L(\hat x, \hat\partial)$ is scaled according to
\be
L\mapsto L_\varepsilon\sim\sum_{k=0}^\infty \varepsilon^{k-2}L^{\rm int}_k\,,
\ee
where $L^{\rm int}_k$ are second-order differential operators with homogeneous 
symbols. 
Next, we expand the scaled heat kernel in $M_{\rm int}$, which we denote by
$U^{\rm int}_\varepsilon$ in a power series in $\varepsilon$
\be
U^{\rm int}_\varepsilon\sim \sum_{k=0}^\infty \varepsilon^{2-n+k}
U^{\rm int}_k\,,
\ee
and substitute into the scaled version of the heat equation.
By equating the like powers of $\varepsilon$ we
get an infinite set of recursive differential equations determining all the
coefficients $U^{\rm int}_k$.

The leading order operator $L^{\rm int}_0$ is an operator
with constant coefficients determined by the leading symbol
\be
L^{\rm int}_0=H(x_0,-i\partial)\,.
\ee
The leading-order heat kernel $U^{\rm int}_0$ can be easily obtained by
the Fourier transform
\be
U^{\rm int}_{0}(t;x,x')=\int\limits_{\RR^n}\frac{d\xi}{(2\pi)^n}
e^{i\xi (x-x')-tH(x_0,\xi)}\,,
\ee
where $\xi(x-x')=\xi_\mu(x^\mu-x'^\mu)$.

The higher-order coefficients $U^{\rm int}_k$, $k\ge 1$, 
are determined from the recursive equations
\be
(\partial_t+L^{\rm int}_0)U^{\rm int}_{k}=-\sum_{j=1}^k L^{\rm int}_j 
U^{\rm int}_{k-j}\,,
\ee
with the initial condition
\be
U^{\rm int}_k(0;x,x')=0\,.
\ee

This expansion is nothing but the decomposition of
the heat kernel into the homogeneous parts with respect to the variables
$(x-x_0), (x'-x_0),$ and $\sqrt{t}$. That is,
\be
U^{\rm int}_k(t;x,x')
=t^{(k-n)/2}U^{\rm int}_k\left(1;
x_0+\frac{(x-x_0)}{\sqrt{t}},x_0+\frac{(x'-x_0)}{\sqrt{t}}\right)\,.
\ee
In particular, the heat kernel diagonal at the point $x_0$
scales by
\be
U^{\rm int}_k(t;x_0,x_0)
=t^{(k-n)/2}U^{\rm int}_k\left(1;x_0,x_0\right)\,.
\label{ukhom}
\ee

To compute the contribution of these coefficients to the trace of the
heat kernel we need to compute the integral of the diagonal of the heat
kernel  $U^{\rm int}(t;x,x)$ over the interior part
of the manifold $M_{\rm int}$. 
By using the homogeneity property (\ref{ukhom}) we obtain
\bea
\int\limits_{M_{\rm int}}d x\; \tr_S U^{\rm int}(t;x,x)
&\sim&
\sum_{k=0}^\infty t^{(k-n)/2} 
\int\limits_{M_{\rm int }}dx\; 
\tr_S U^{\rm int}_k\left(1;x,x\right)\,.
\eea
Next, we take the limit as $\delta\to 0$.  Then the integrals over
the interior part $M_{\rm int}$ become the integrals over the whole
manifold $M$ and give all the interior coefficients $a_k(L)$ in the
global heat kernel coefficients $A_k(L)$.

Instead of this rigorous procedure, we present below a 
pragmatic formal  approach that enables one to compute all interior
coefficients in a much easier and compact form. Of course, both
approaches are equivalent and give the same answers.

First, we present the heat kernel diagonal for the operator 
$L=\bar{D}\,D$
in the form
\begin{equation}
U^{\rm int}(t;x,x) =\int\limits_{\RR^n} {d\xi\over (2\pi)^n} e^{-i\xi
  x}\exp(-t\,\bar{D}\,D) e^{i\xi x} \,,
\end{equation}
where $\xi x=\xi_\mu x^\mu$, which can be transformed to
\begin{equation}
U^{\rm int}(t;x,x)=\int\limits_{\RR^n} {d\xi\over (2\pi)^n}
\exp\left[-t\left(H+K+\,\bar{D}\,D\right)\right]\cdot \II\,,
\end{equation}
where $H=[\Gamma(\xi)]^2$ is the leading symbol of the operator 
$\bar{D}\,D$,
and $K$ is a first-order self-adjoint operator defined by
\be
K=-\Gamma(\xi)D-\,\bar{D}\,\Gamma(\xi)\,.
\ee
Here the
operators in the exponent act on the unity matrix $\II$ from the left.

By changing the integration variable $\xi\to t^{-1/2}\xi$ we obtain
\begin{equation}
U^{\rm int}(t;x,x)=(4\pi t)^{-n/2}\int\limits_{\RR^n} {d\xi\over \pi^{n/2}} 
\exp\left(-H-\sqrt{t}\,K-t\,\bar{D}\,D\right)\cdot \II\,.
\end{equation}
Now, the coefficients of the asymptotic expansion of this integral in
powers of $t^{1/2}$  as $t\to 0$ determine the interior heat kernel
coefficients $a_k(L)$ via
\be
\tr_S U^{\rm int}(t;x,x)
\sim (4\pi)^{-n/2}\sum_{k=0}^\infty t^{(k-n)/2}a_k(L)\,.
\ee

By using the Volterra series
\begin{eqnarray}
\exp(A+B)&=&e^A+\sum\limits_{k=1}^\infty
\int\limits_0^1 d\tau_k \int\limits_0^{\tau_k}d\tau_{k-1}\cdots 
\int\limits_0^{\tau_2} d\tau_1\times
\nonumber\\[10pt]
&&
\times\, e^{(1-\tau_{k})A} Be^{(\tau_k-\tau_{k-1}) A} \cdots
e^{(\tau_2-\tau_1)A} Be^{\tau_1 A}\,,
\end{eqnarray}
we get
\begin{eqnarray}
&&\exp\left(-H-\sqrt{t}\,K
-t\,\bar{D}\,D\right)
=e^{-H}-
t^{1/2}\int\limits_0^1 d\tau_1 e^{-(1-\tau_1)H}K e^{-\tau_1 H}
\nonumber\\
&&
\qquad\qquad\qquad
+\,t\Biggl[ \int\limits_0^1d\tau_2\int\limits_0^{\tau_2}d\tau_1
  e^{-(1-\tau_2)H} K e^{-(\tau_2-\tau_1)H}Ke^{-\tau_1 H}-
\nonumber\\
&&
\qquad\qquad\qquad
\hphantom{+\,t\Biggl[}
  -\int\limits_0^1 d\tau_1 e^{-(1-\tau_1)H}
\,\bar{D}\,D e^{-\tau_1 H} \Biggr]+
O(t^2)\,.
\end{eqnarray}

Now, since $K$ is linear in $\xi$ the term proportional to $t^{1/2}$
vanishes after integration over $\xi$.
Thus, we obtain the first three interior
coefficients of the asymptotic expansion of the
heat kernel diagonal in the form
\begin{eqnarray}
a_0(L) &=&\int\limits_{\RR^n}{d\xi\over \pi^{n/2}}\,
\tr_S \,e^{-H}\,,
\\
a_1(L)&=& 0\,,
\\
a_2(L)&=&\int\limits_{\RR^n}{d\xi\over \pi^{n/2}}\,
\tr_S
\Biggl[
\int\limits_0^1d\tau_2\int\limits_0^{\tau_2}d\tau_1 e^{-(1-\tau_2)H}
K e^{-(\tau_2-\tau_1)H}Ke^{-\tau_1 H}-
\nonumber\\&&
         \hphantom{\int\limits_{\RR^n}{d\xi\over \pi^{n/2}}\,\Biggl[}
-\int\limits_0^1 d\tau_1 e^{-(1-\tau_1)H}
\,\bar{D}\,D e^{-\tau_1 H}
\Biggr] \,.
\end{eqnarray}
 
\subsection{Boundary Coefficients}

On manifolds with boundary, as far as we know, the coefficients $A_k$
have not been  studied at all, so, even $A_1$ is not known. In the
present paper we are going to compute the coefficient $A_1$ on manifolds
with boundary for the operators $\bar DD$ and $D\bar D$.  The
coefficient $A_0$ is, of course, the same as for the manifolds without
boundary. We will follow the general framework for computation of the
heat kernel asymptotics outlined in \cite{avramidi04c,avramidi99b}.

The procedures for the resolvent and the heat kernel are very similar.
One can, of course, use either of them. We will describe below the 
construction of the heat kernel.

The main idea can be described as follows. Recall that we decomposed the
manifold into a neighborhood of the boundary $M_{\rm bnd}$ and the
interior part $M_{\rm int}$. We can use now different approximations for
the heat kernel in different domains. Strictly speaking one has to use 
`smooth characteristic functions' of those domains (partition of unity)
to glue them together in a smooth way. Then, one has to control the
order of the remainder terms in the limit $t\to 0^+$ and their
dependence on $\delta$ (the size of the boundary layer). However, since
we are only interested in the trace of the heat kernel, this is not
needed here and we will not worry about such subtle details. We can
compute the asymptotic expansion as $t\to 0$ of the corresponding
integrals in each domain and then take the limit $\delta\to 0$.

The origin of the boundary terms in the heat trace asymptotics can be
explained as follows. The heat kernel of an elliptic boundary value
problem in $M_{\rm bnd}$ has exponentially small terms like
$\exp(-r^2/t)$ as $t\to 0$. These terms do not contribute to the
asymptotic expansion of the diagonal of the heat kernel as $t\to 0$.
However, they behave like distributions near the boundary (recall that
$r>0$ inside the manifold and $r=0$ on the boundary). Therefore, the
integral over $M_{\rm bnd}$, more precisely, the limit $\lim_{\delta\to
0}\int\limits_{\partial M}d\hat x\int\limits_0^\delta dr(\dots)$ does 
contribute to the
asymptotic expansion of the trace of the heat kernel with coefficients
in form of integrals over the boundary. It is this phenomenon that leads
to the boundary terms in the global heat invariants.

The heat kernel in the boundary layer $M_{\rm bnd}$ is constructed as
follows. We fix a point $\hat x_0\in\partial M$ on the boundary and choose
coordinates as described above in section 2.2. Let $\varepsilon>0$ be a 
positive real parameter. We use it as a scaling parameter; at the very
end of the calculation it will be set to $1$.  Now we scale the
coordinates according to 
\be 
\hat x^j\mapsto \hat x^j_0+\varepsilon(\hat
x^j-\hat x^j_0)\,,\qquad  \hat x'^j\mapsto \hat x^j_0+\varepsilon(\hat
x'^j-\hat x^j_0)\,, \ee \be r\mapsto \varepsilon r\,, \qquad 
r'\mapsto\varepsilon r'\,,\qquad t\mapsto\varepsilon^2 t\,. 
\ee 
The
differential operators are scaled correspondingly by 
\be
\hat \partial_j
\mapsto\frac{1}{\varepsilon}\hat\partial_j\,,\qquad \partial_r
\mapsto\frac{1}{\varepsilon}\partial_r\,,\qquad \partial_t
\mapsto\frac{1}{\varepsilon^2}\partial_t\,. 
\ee  

Let $L(r,\hat x,
\partial_r,\hat\partial)$ be the operator under  consideration. The
scaled operator, which we denoted by $L_\varepsilon$, has the following
formal power series expansion  in $\varepsilon$ 
\be 
L\mapsto
L_\varepsilon\sim\sum_{k=0}^\infty \varepsilon^{k-2}L^{\rm bnd}_k\,, 
\ee 
where
$L_k$ 
are second-order differential operators with homogeneous  symbols.
The leading order operator is determined by the leading symbol 
\be
L^{\rm bnd}_0=H(0,\hat x_0,-i\partial_r,-i\hat\partial)\,. 
\ee This is a
differential operator with constant coefficients.  

Next, we expand the
scaled heat kernel in $M_{\rm bnd}$, which we denote by $U^{\rm
bnd}_\varepsilon$ in a power series in $\varepsilon$ 
\be 
U^{\rm
bnd}_\varepsilon\sim \sum_{k=0}^\infty \varepsilon^{2-n+k} U^{\rm
bnd}_k\,, 
\ee 
and substitute into the scaled version of the heat
equation and the  boundary conditions.   By equating the like powers of
$\varepsilon$ we get an infinite set of recursive differential equations
determining all the coefficients $U^{\rm bnd}_k$.  

The leading-order
heat kernel $U^{\rm bnd}_0$ is determined by the equation 
\be
(\partial_t+L^{\rm bnd}_0)U^{\rm bnd}_{0}=0 
\ee 
with the initial condition 
\be
U^{\rm bnd}_0(0;r,\hat x,r',\hat x')=\delta(r-r')\delta(\hat x,\hat
x')\,, 
\ee 
the boundary condition 
\be 
U^{\rm bnd}_0(t;0,\hat x,r',\hat
x')= U^{\rm bnd}_0(t;r,\hat x,0,\hat x')=0\,, 
\ee 
and the asymptotic
condition 
\be 
\lim_{r\to\infty}U^{\rm bnd}_0(t;r,\hat x,r',\hat x')
=\lim_{r'\to\infty}U^{\rm bnd}_0(t;r,\hat x,r',\hat x')=0\,. 
\ee 

The
higher-order coefficients $U^{\rm bnd}_k$, $k\ge 1$,  are determined
from the recursive equations 
\be 
(\partial_t+L^{\rm bnd}_0)U^{\rm
bnd}_{k}=-\sum_{j=1}^k L^{\rm bnd}_j U^{\rm bnd}_{k-j}\,, 
\ee
 with the initial
condition \be U^{\rm bnd}_k(0;r,\hat x,r',\hat x')=0\,, 
\ee 
the boundary
condition \be U^{\rm bnd}_k(t;0,\hat x,r',\hat x')= U^{\rm
bnd}_k(t;r,\hat x,0,\hat x')=0\,, \ee and the asymptotic condition 
\be
\lim_{r\to\infty}U^{\rm bnd}_0(t;r,\hat x,r',\hat x')
=\lim_{r'\to\infty}U^{\rm bnd}_0(t;r,\hat x,r',\hat x')=0\,. 
\ee  

This
expansion is nothing but the decomposition of the heat kernel into the
homogeneous parts with respect to the variables $(\hat x-\hat x_0),
(\hat x'-\hat x_0), r,r'$ and $\sqrt{t}$. That is, 
\be
 U^{\rm
bnd}_k(t;r,\hat x,r',\hat x') =t^{(k-n)/2}U^{\rm
bnd}_k\left(1;\frac{r}{\sqrt{t}}, \hat x_0+\frac{(\hat x-\hat
x_0)}{\sqrt{t}},\frac{r'}{\sqrt{t}}, \hat x_0+\frac{(\hat x'-\hat
x_0)}{\sqrt{t}}\right)\,.
 \ee 
 In particular, the heat kernel diagonal at
the point $(r,\hat x_0)$ scales by 
\be 
U^{\rm bnd}_k(t;r,\hat x_0,r,\hat
x_0) =t^{(k-n)/2}U^{\rm bnd}_k\left(1;\frac{r}{\sqrt{t}}, \hat
x_0,\frac{r}{\sqrt{t}}, \hat x_0\right)\,. 
\label{ukhombnd}
\ee  

To compute the
contribution of these coefficients to the trace of the heat kernel we
need to compute the integral of the diagonal of the heat kernel  $U^{\rm
bnd}(t;r,\hat x,r,\hat x)$ over the boundary layer $M_{\rm bnd}$. This
heat kernel diagonal can be decomposed as the sum of two terms, the
first coming from the standard interior heat kernel on manifolds without
boundary (that does not satisfy the boundary conditions) and the second
`compensating' part, which is the crucial  boundary part and whose role
is to make the heat kernel to satisfy the boundary conditions (for more
details see \cite{avramidi04c}). The integral of the `boundary' part
over the boundary layer in the limit when the size of the boundary layer
goes to zero produces the boundary contributions $b_k(L)$ to the global
heat kernel coefficients $A_k(L)$.  

By using the homogeneity property (\ref{ukhombnd}) we
obtain 
\bea 
\int\limits_{M_{\rm bnd}}d x\; \tr_S U^{\rm bnd}(t;x,x)
&=&\int\limits_{\partial M}d\hat x  \int\limits_0^\delta dr\; \tr_S  U^{\rm
bnd}(t;r,\hat x,r,\hat x) \nonumber\\ 
&\sim& \sum_{k=0}^\infty
t^{(k-n)/2}  \int\limits_{\partial M}d\hat x  \int\limits_0^\delta dr\; 
\tr_S  U^{\rm
bnd}_k\left(1;\frac{r}{\sqrt{t}},\hat x,\frac{r}{\sqrt{t}},\hat x\right)
\nonumber\\ 
&\sim& \sum_{k=0}^\infty t^{(k-n+1)/2}  \int\limits_{\partial
M}d\hat x  \int\limits_0^{\delta/\sqrt{t}} du\; \tr_S  U^{\rm
bnd}_k\left(1;u,\hat x,u,\hat x\right) 
\nonumber \\ 
\eea
 where
$u=r/\sqrt{t}$. Notice the appearance of the extra power of $\sqrt{t}$ in the
asymptotic expansion. Of course, if one takes the limit $\lim_{\delta\to
0}$ for a finite $t$, then all these integrals vanish. However, if one
takes  the limit $\lim_{t\to 0}$ first for a finite $\delta$, and then
the limit $\lim_{\delta\to 0}$, then one gets finite answers for the
boundary coefficients $b_k(L)$.

\subsubsection{Leading-Order Heat Kernel}

To compute the coefficient $A_1$ we just need the  leading-order heat
kernel $U^{\rm bnd}_0$. We will, in fact, be working in the tangent
space $\RR_+\times T_{\hat x_0}\partial M$ 
at a point $\hat x_0$ on the boundary
and reduce our problem to a
problem on the half-line.  The operator $L^{\rm bnd}_0$ acts on square
integrable sections of the vector bundle $\mathcal{S}[\frac{1}{2}]$ in a
neighborhood of the point $\hat x_0$. We extend the operator appropriately to
the space $L^2(\mathcal{S}[\frac{1}{2}],
\RR_+,\RR^{n-1}, dr\,d\hat x)$ so that it coincides with the initial
operator in the neighborhood of the point $\hat x_0$. When computing the 
trace below we set $\hat x_0=\hat x=\hat x'$.

By using the Laplace transform in the variable $t$ and the Fourier
transform in the boundary coordinates $\hat x$
\be
U^{\rm bnd}_0(t;r,\hat x,r',\hat x')=
\frac{1}{2\pi i}\int\limits_{w-i\infty}^{w+i\infty} d\lambda\;
\int\limits_{\RR^{n-1}}\frac{d\hat\xi}{(2\pi)^{n-1}}
e^{-t\lambda+i\hat\xi(\hat x-\hat x')}\,
F(\lambda,,r,r',\hat\xi)\,,
\ee
we obtain an ordinary differential equation
\be
\left(-A^2\partial_r^2-iB\partial_r+C^2-\lambda\II\right)
F(\lambda,r,r',\hat\xi)=\II\delta(r-r')
\label{odef}
\ee
where the matrices $A$, $B$ and $C$ are 
defined in (\ref{matac}), (\ref{matb}),
and are frozen at the point $\hat x_0$ (they are constant for the 
purpose of this calculation),
with the boundary condition
\be
F(\lambda,0,r',\hat\xi)=F(\lambda,r,0,\hat\xi)=0
\ee
the asymptotic condition
\be
\lim_{r\to\infty}F(\lambda,r,r',\hat\xi)
=\lim_{r'\to\infty}F(\lambda,r,r',\hat\xi)=0\,,
\ee
and the self-adjointness condition
\be
\overline{F(\lambda,r,r',\hat\xi)}
=F(\bar\lambda,r',r,\hat\xi)\,.
\label{asympf}
\ee
It is easy to see that $F$ is a homogeneous function
\be
F\left(\frac{\lambda}{t},
\sqrt{t}r,\sqrt{t}r',\frac{\hat \xi}{\sqrt{t}},\right)
=t^{1/2}F\left(\lambda,r,r',\hat\xi\right)\,.
\label{hom}
\ee

We decompose the Green function in two parts,
\be
F=F_\infty+F_B\,,
\ee
where $F_\infty$ is the part that is valid for the
whole real line and $F_B$ is the compensating term. The part $F_\infty$
can be easily obtained by the
Fourier transform; it has the form
\be
F_\infty(\lambda,r,r',\hat\xi)
=\Phi(\lambda,r-r',\hat\xi)\,,
\ee
where $\Phi(\lambda,r,\hat\xi)$ is defined in (\ref{kdef}).
It is not smooth at the diagonal $r=r'$ and is
responsible for the appearance of the delta-function $\delta(r-r')$ on
the right-hand side of the eq. (\ref{odef}). 

The corresponding part of the leading heat kernel is then easily computed
to be
\be
U_{0,\infty}^{\rm bnd}{}(t;x,x')
=\int\limits_{\RR^n}\frac{d\xi}{(2\pi)^n} e^{i\xi(x-x')-tH(x_0,\xi)}\,,
\ee
where $x_0=(0,\hat x_0)$.
This part does not contribute to the asymptotics of the  trace of the
heat kernel in the limit $\delta\to 0$. By rescaling $\xi\mapsto
\xi/\sqrt{t}$ we obtain
\be
\int\limits_{M_{\rm bnd}}dx\;\tr_S\,U_{0,\infty}^{\rm bnd}{}(t;x,x)
=(4\pi t)^{-n/2}\int\limits_{M_{\rm bnd}}dx\;
\int\limits_{\RR^n}\frac{d\xi}{\pi^{n/2}}
\tr_S e^{-H(x,\xi)}\,,
\ee
and in the limit $\delta\to 0$ this integral vanishes.

However, $F_\infty$  does not  satisfy the boundary conditions. The role
of the boundary part, $F_B$, is exactly to guarantee that $F$
satisfies the boundary
conditions. The function $F_B$ is smooth at the diagonal $r=r'$. 
It can be presented in the following form
\be
F_B(\lambda,r,r',\hat\xi)
=-\Phi(\lambda,r,\hat\xi)
[\Phi_0(\lambda,\hat\xi)]^{-1}
\Phi(\lambda,-r',\hat\xi)\,.
\label{bndgr}
\ee

\subsubsection{The Coefficient $A_1$}

The coefficient $A_1$ is a pure boundary coefficient that is computed by
integrating the boundary part $U_{0,B}^{\rm bnd}$ of the heat kernel. We
have
\bea
&&\int\limits_{M_{\rm bnd}}dx\;\tr_S\,U_{0,B}^{\rm bnd}{}(t;x,x)
\\
&&\qquad\qquad
=\int\limits_{\partial M} d\hat x\,
\int\limits_0^\delta dr\, 
\int\limits_{\RR^{n-1}}\frac{d\hat \xi}{(2\pi)^{n-1}} 
\int\limits_{w-i\infty}^{w+i\infty}
\frac{d\lambda}{2\pi i}e^{-t\lambda}
\tr_S\,F_B(\lambda,r,r,\hat\xi)\,.
\nonumber
\eea

Now, by rescaling the variables 
\be
\lambda\mapsto \frac{\lambda}{t},\qquad
r\mapsto \sqrt{t}r,\qquad
\hat\xi\mapsto \frac{\hat\xi}{\sqrt{t}}
\ee
and using the homogeneity property (\ref{hom})
we obtain
\bea
&&\int\limits_{M_{\rm bnd}}dx\;\tr_S\,U_{0,B}^{\rm bnd}{}(t;x,x)
\\
&&
=t^{(1-n)/2}
\int\limits_{\partial M} d\hat x\,
\int\limits_{\RR^{n-1}}\frac{d\hat \xi}{(2\pi)^{n-1}} 
\int\limits_0^{\delta/\sqrt{t}} dr\, 
\int\limits_{w-i\infty}^{w+i\infty}
\frac{d\lambda}{2\pi i}e^{-\lambda}
\tr_S\,
F_B\left(\lambda,r,r,\hat\xi\right)\,.
\nonumber
\eea

Therefore, the coefficient $A_1$ is given by
\be
A_1=
2\sqrt{\pi}
\int\limits_{\partial M} d\hat x\,
\int\limits_{\RR^{n-1}}\frac{d\hat \xi}{\pi^{(n-1)/2}} 
\int\limits_0^{\infty} dr\, 
\int\limits_{w-i\infty}^{w+i\infty}
\frac{d\lambda}{2\pi i}e^{-\lambda}
\tr_S\,F_B\left(\lambda,r,r,\hat\xi\right)\,.
\ee
Thus, finally, by using eq. (\ref{bndgr}), 
eliminating the odd functions of $\hat\xi$ (since the integrals
of them vanish), using the property (\ref{phixi}) of the function $\Phi$ 
 and extending the integration over $r$ from $-\infty$ 
to $+\infty$ (since the integrand is an even function)
we obtain
\be
A_1=
\int\limits_{\partial M} d\hat x\, 
\int\limits_{\RR^{n-1}}\frac{d\hat \xi}{\pi^{(n-1)/2}} 
\Psi_1(\hat\xi)
\ee
where
\bea
\Psi_1(\hat\xi)&=&
-\frac{\sqrt{\pi}}{2}
\int\limits_{-\infty}^{\infty} dr\, 
\int\limits_{w-i\infty}^{w+i\infty}
\frac{d\lambda}{2\pi i}e^{-\lambda}\,\tr_S\,
[\Phi_0(\lambda,\hat\xi)]^{-1}
\\
&&\times
\Bigl\{\Phi(\lambda,r,\hat\xi)
\Phi(\lambda,-r,\hat\xi)
+
\Phi(\lambda,-r,\hat\xi)
\Phi(\lambda,r,\hat\xi)\Bigr\}
\nonumber
\,.
\eea
Recall that $w$ is a sufficiently large negative constant. 

Now, using eq. (\ref{73}) and integrating over $r$ we obtain
finally
\bea
\Psi_1(\xi)&=&
-\sqrt{\pi}
\int\limits_{w-i\infty}^{w+i\infty}
\frac{d\lambda}{2\pi i}\,e^{-\lambda}\,
\tr_S\,[\Phi_0(\lambda,\hat\xi)]^{-1}
\frac{\partial}{\partial\lambda}
\Phi_0(\lambda,\hat\xi)\,
\nonumber\\
&=&
-\sqrt{\pi}
\int\limits_{w-i\infty}^{w+i\infty}
\frac{d\lambda}{2\pi i}\,e^{-\lambda}\,
\frac{\partial}{\partial\lambda}
\log
\det\,[\Phi_0(\lambda,\hat\xi)]\,.
\eea

Thus, the problem is now reduced to the computation of the integral
over $\lambda$. This is not at all trivial because of the
presence  of two non-commuting matrices, essentially,
$A^{-1}(AC+CA)A^{-1}$ and $A^{-1}(C^2-\lambda\II)A^{-1}$,  where the
matrices $A=\Gamma^r(\hat x)$ and   $C=\Gamma^j(\hat x)\hat\xi_j$ are
defined by (\ref{matac}). We will report on this problem in a future
work. Here let us just mention that in the particular case when
$B=AC+CA=0$ (for example, this is so in the case of the original Dirac
operator) we get
\be
\tr_S\,[\Phi_0(\lambda,\hat\xi)]^{-1}
\frac{\partial}{\partial\lambda}\Phi_0(\lambda,\hat\xi)
=\frac{1}{2}\tr_S\,(C^2-\lambda\II)^{-1}\,,
\ee
and, therefore, one can compute the integral over $\lambda$ 
to obtain
\be
A_1=
-\frac{\sqrt{\pi}}{2}
\int\limits_{\partial M} d\hat x\,
\int\limits_{\RR^{n-1}}\frac{d\hat \xi}{\pi^{(n-1)/2}}\,
\tr_S\,
e^{-[C(\hat x,\hat\xi)]^2}
\,.
\ee
Of course, for Laplace type operators, when $[C(\hat
x,\hat \xi)]^2=\II g^{ij}(\hat x)\hat
\xi_i\hat\xi_j$, the integral can be computed explicitly, which  gives
the induced Riemannian volume of the boundary,
$A_1=-(\sqrt{\pi}/2)\,N\,\vol(\partial M)$, and coincides with the
standard result for Dirichlet Laplacian \cite{gilkey95}.

\section*{Acknowledgements}

I would like to thank the organizers of the conference ``Boundary Value
Problems and Friends'', Bernhelm Booss-Bavnbek, Slawomir Klimek and
Ryszard Nest, for their kind invitation to present this work. It was a
pleasure to contribute to this  special issue dedicated to Krzysztof
Wojciechowski. We all wish him soon recovery and best of luck in the
future.

\end{document}